\documentclass[12pt,preprint]{aastex}

\shorttitle{Accretion onto Black Hole - Parameter Space}
\shortauthors{Manickam \& Chakrabarti}

\begin{document}

\title{Accretion onto a Black Hole in the Presence of Bremsstrahlung Cooling - 
Parameter Space Study} 

\author{Sivakumar G. Manickam and Sandip K. Chakrabarti} 
\affil{S. N. Bose National Centre for Basic Sciences, JD block, Sector III, Salt lake, \\ Kolkata 700 098, India} 
\email{sivman@boson.bose.res.in and chakraba@boson.bose.res.in}

\begin{abstract}
The dynamics of accretion onto a schwarzschild black hole is studied using Paczynski-Wiita
pseudo newtonian potential. Steady state solution of the flow equations is obtained using thin
disc approximation, including the effect of bremsstrahlung cooling in the energy equation. The
topology of transonic solutions are got and the conditions for shock formation (Rankine-Hugoniot
conditions) are checked. Shock and no-shock regions in the parameter space of accretion rate and
specific angular momentum are identified. We motivate the time-dependent numerical simulation
of the flow, as a candidate for explaining the quasi periodic oscillations observed in black hole
candidates.
\end{abstract}


\section{Introduction}

The parameter space of shock formation in adiabatic transonic accretion discs was first discussed  
in Chakrabarti(1989). Though time-dependent numerical simulation of transonic accretion 
disc in the presence of cooling was dealt in Molteni, Sponholz \& Chakrabarti (1996, hereafter MSC96),
the detailed study of parameter space was reserved for the future. In this paper we describe the 
procedure for obtaining the flow topology in the presence of bremsstrahlung cooling and obtain
the dependence of topology and shock formation on the parameters, accretion rate and specific angular
momentum. 

\section{Flow equations} 

The equations which govern the flow are the basic conservation equations of mass, momentum and
energy. The pseudo newtonian potential of Paczynski-Wiita (1980) is used, which mimics the gravitational
field around a schwarzschild black hole to a sufficient accuracy. The effect of bremsstrahlung cooling (Lang 1980)
in electron-proton plasma is included in the energy equation. ${\bf v}(v_r, v_{\phi}, v_z), \rho$ and 
$p$ are the velocity, density and pressure respectively.

\noindent Continuity equation:

\begin{equation}
{{{\partial \rho}\over{\partial t}} +{ \nabla .{({\rho}{\bf v})}} = 0 } 
\end{equation}

\noindent Euler equation:

\begin{equation}
{{\partial \bf v} \over {\partial t}} + (\bf v. \nabla ) \bf v = - {\nabla p \over \rho} - \nabla g
\end{equation}

\noindent where $ g = -{GM \over (r-r_g)} $ is the Paczynski-Wiita potential, $G$ is the gravitational
constant, M is the mass of the black hole, $r_g={2GM \over c^2}$ is the schwarzschild radius
and $c$ is the velocity of light.

\noindent Energy equation:

\begin{equation}
\nabla . ( \rho \epsilon \bf v) + \Lambda - \Gamma = - {\partial \over {\partial t}} (\rho \epsilon)
\end{equation}

\noindent where $ \epsilon = {1 \over 2} {v_r}^2 + U + {p \over \rho} + g + {1 \over 2} {v_ \phi}^2 $ 
is the specific energy, $U={p \over \rho (\gamma-1)}$ is the thermal energy,
$\gamma$ is the adiabatic index,
$ \Lambda = 1.43 \times 10^{-27} {\rho ^2 \over m_p ^2} T^{1/2} g_f $ is the expression 
for bremsstrahlung cooling,
$T$ is the temperature, $m_p$ is the mass
of proton, $g_f$ is the gaunt factor and $\Gamma$ is the heating term. 

\noindent Thin disc approximation:

\noindent We use cylindrical polar coordinates $(r,\phi,z)$ for the axisymmetric flow. Because of axisymmetry and
thinness of the disc the following assumptions are made. 

$ \rho(\phi^0, z^0) $ \ \ \ \ \ i.e. $\rho$ is not a function of $\phi$ and $z$ 

$ {v_r}(\phi^0, z^0) $ 

$ {v_ \phi}(\phi^0, z^0) $

$ {v_z}(r^0, \phi^0, z^0) $

$ p(r) $ \ \ \ \ \ i.e. $p$ is a function of r only  

$ g(r) $

$ \epsilon(\phi^0) $ 

\noindent By the help of these assumptions it is possible to write the steady state equations in the
following form.

\begin{equation}
{{\psi ^ \prime - a^2 /r +j{g_f}{2 \over 3} {\rho \over v {m_p}^2} T^{1/2}} \over a^2 /v -v}
= {dv \over dr} \ ; \ \psi = -{GM \over (r-r_g)} + {1 \over 2} {v_ \phi}^2 
\end{equation}

\begin{equation}
{1 \over \rho} {d \rho \over dr} + {1 \over v} {dv \over dr} + {1 \over r} = 0 
\end{equation}

\begin{equation}
v{dv \over dr} + {2a \over \gamma}{da \over dr}+{a^2 \over \rho \gamma}{d\rho \over dr}+\psi ^ \prime=0 
\end{equation}

\noindent where $^\prime$ denotes the derivative with respect to r, $j=1.43 \times 10^{-27}$ in cgs units and the 
polytropic relation $p=K \rho ^ \gamma$ is 
used to obtain the sound speed $a=\sqrt
{\partial p \over 
\partial \rho}=\sqrt{\gamma p \over \rho}$. 
These equations are solved using fourth-order Runge-Kutta method (Press et al. 1992).  

\section{Sonic point analysis}                    

The flow can pass through a point where denominator in the expression for ${dv \over dr}$
becomes zero. If it happens that the numerator also becomes zero, then 
${dv \over dr}$ can be finite. We call such a point a critical point. 
In this problem the critical point is same as sonic point. 
At critical point, using l' Hospital's rule, we get (for $\gamma={5\over 3}$),   

\begin{equation} {dv \over dr} = {-B \pm {\sqrt{(B^2-4AC)}} \over 2A} \end{equation} 

\begin{equation} A = {8 \over 3} \end{equation} 

\begin{equation}
B = -{1\over3}{v\over r}-{4\over 3} \zeta \rho a^{2\alpha-2} (1+{\alpha\over 3}) +{5\over 3}{\psi^\prime \over v} 
\end{equation}

\begin{equation}
C = {5\over 3}{\psi ^ \prime \over r}+ \psi ^ {\prime \prime} -{2\over 3}\zeta {a^{2\alpha-1}\over r}
       \rho (1-\alpha)-{10\over9}\alpha\rho\zeta a^{2\alpha-3}
       \psi ^\prime
\end{equation}

\begin{equation}
\zeta = jg_f{1\over m_p^2} ({\mu m_p \over \gamma k})^\alpha
\end{equation}

\noindent where $\alpha=0.5$, $\mu=0.5$ and $k$ is the boltzmann constant.  
Now starting from the critical point we obtain 
the transonic solution topologies.

\section{Numerical scheme}

Fluid dynamical problems are inherently sensitive to the boundary conditions. We formulate
the problem as follows. 
The variables which are functions of space and time are
${\bf v}, \rho, p$ \  or \ ${\bf v}, \rho, a$. 
The parameters are accretion rate, specific angular momentum($\lambda$), $\gamma$,  
cooling process and viscosity. We consider
inviscous flow, set
$\gamma = 5/3$ and the mass of the black hole $M$ is chosen as $10^8 M_{\odot}$.
So the free parameters are accretion rate and $\lambda$. 
The location of the critical points, $r_{c1}, r_{c2}$, is the
freedom we have, subjected to the  
constraint that shock conditions should be satisfied. 
The 
natural constraint of stability is likely to decide the uniqueness of the solution. 
The algorithm for obtaining the numerical solution is,

1. Choose accretion rate and $\lambda$

2. Choose $r_{c1}$ and $r_{c2}$ range and find corresponding $a_{c1}$ and $a_{c2}$  

3. Obtain solution topology by using fourth-order Runge-Kutta method

4. Check the shock conditions 

\noindent The typical components of the topology are shown in fig.1, where mach number is plotted
as a function of the radial distance. 
Fig.2 and 3 shows the solution topologies for chosen parameter values.
Rankine-Hugoniot shock conditions (Landau \& Lifshitz 1984) basically ensure that mass, momentum and energy
are conserved in spite of a discontinuity. For infinitesimally thin and non-dissipative shock,
the shock conditions take the form,

\begin{equation}
p_1 + \rho _1 {v_1}^2 = p_2 + \rho _2 {v_1}^2, \ \ \ \ \ {1 \over 2} {v_1}^2 + {{a_1}^2 \over \gamma -1} = {1 \over 2} {v_2}^2 + {{a_2}^2 \over \gamma -1}
\end{equation}

\noindent where the suffixes 1 and 2 refer to pre-shock and post-shock quantities.

\section{Parameter space} 

We use the following procedure to obtain the parameter space. 
The accretion rate in eddington units, is varied in the range (.0001, 500) and $\lambda$ in the units
of ${2GM \over c}$, 
is varied in the range $(1.6,2.5)$.  
It is suggested that (Chakrabarti 1996) flows with $\lambda$  
in the range between that of marginally stable and marginally bound orbit would form steady
accretion discs, when self-gravity of the disc is neglected.
For a chosen accretion rate and $\lambda$
we scan the r-axis from 1.5$r_g$ to 1000$r_g$ to find if it can be a critical point.
The critical point can be of X or Alp or plA or lA or x type as shown in fig.1.
X looks like the English alphabet X, Alp like Greek alphabet alpha which opens towards infinity, plA is reflected 
Alp which opens towards inner boundary, lA is plA which doesn't reach the inner boundary when $\lambda$ is high
and x is X which doesn't reach the inner boundary when accretion rate is high.  
For certain parameter values there is an intermediate range of r which cannot be a critical point
(denoted by - in topology column of Table 1). 
Table 1 shows the parameter space. 
We obtain supersonic branches for outer critical points and subsonic branches for inner critical points.
When shock conditions are satisfied the flow makes a transition from supersonic to subsonic branch
and reaches the inner boundary (chosen as 1.5$r_g$) supersonically. When shock conditions are not satisfied
accretion is still possible if X type critical point exists.

\section{Discussion and conclusions}

Of all the different possible branches of accretion for chosen parameter value, the real flow
is likely to choose the branch which is most stable, as perturbations are always present in a real
situation. If the assumption of steady flow is relaxed the flow might still choose a steady 
solution branch if the time scale of change is large. Time-dependent numerical simulation of
the flow (MSC96) shows that flow solution oscillates about the steady state solution
when certain resonance condition is met. The 'perturbations' which occur in
a numerical code and physical perturbations should be related, to increase the faith in numerical results.  
Such numerical studies will be pursued in the future. The oscillation of shock location (MSC96)
would mean the size of post-shock region, which is the source for hard photons (Chakrabarti \& Titarchuk 1995),
is also oscillating. These would result in quasi periodic oscillations as reported in Chakrabarti \& Manickam (2000). 

\acknowledgments 

We thank Indian Space Research Organization for funding the project, Quasi-Periodic
Oscillations in Black Hole Candidates, of which this work forms a part.

\begin{figure}
\plotone{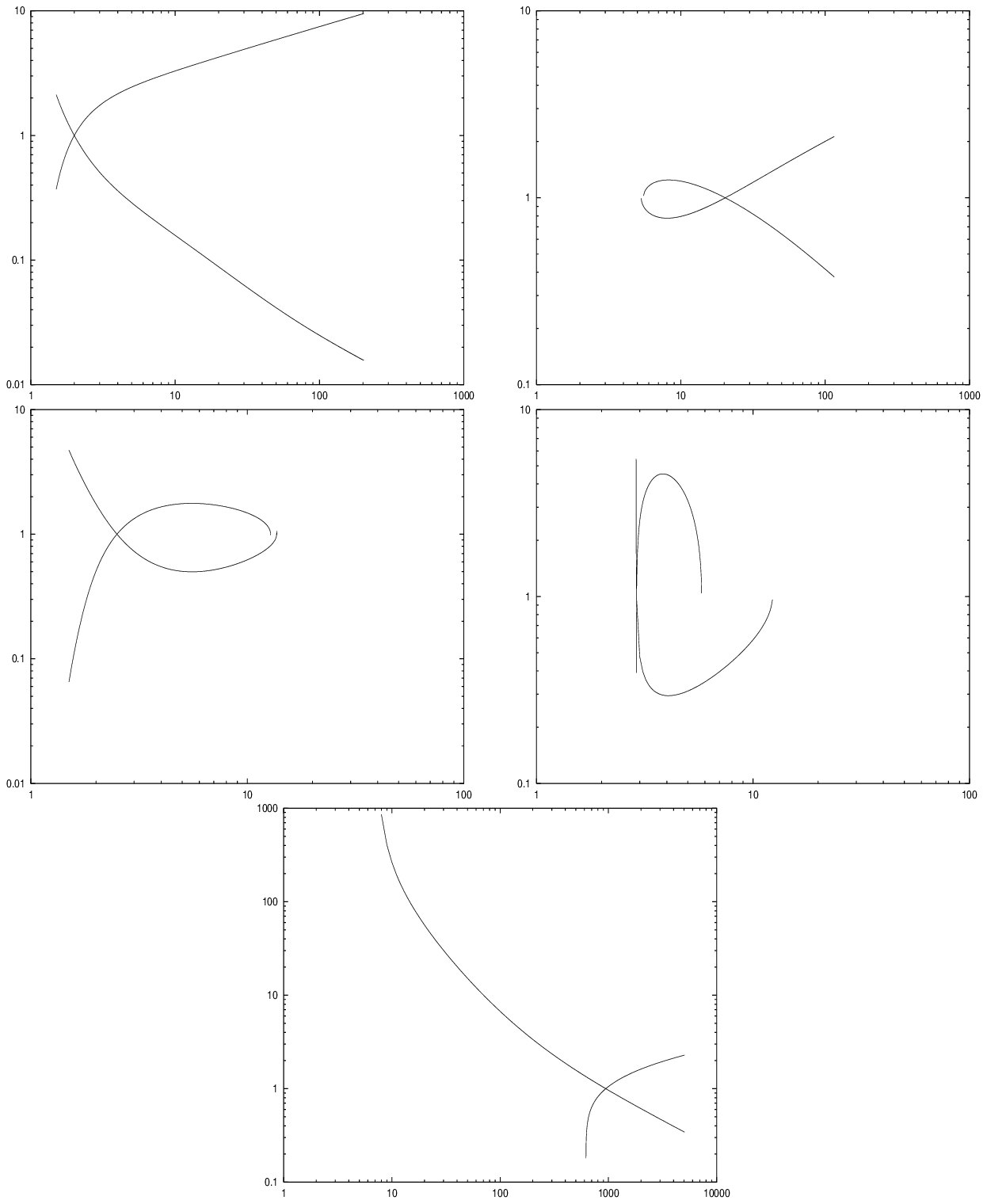}
\caption{Mach number Vs radial distance of X, Alp, plA, lA and x} 
\end{figure}

\begin{figure}
\plotone{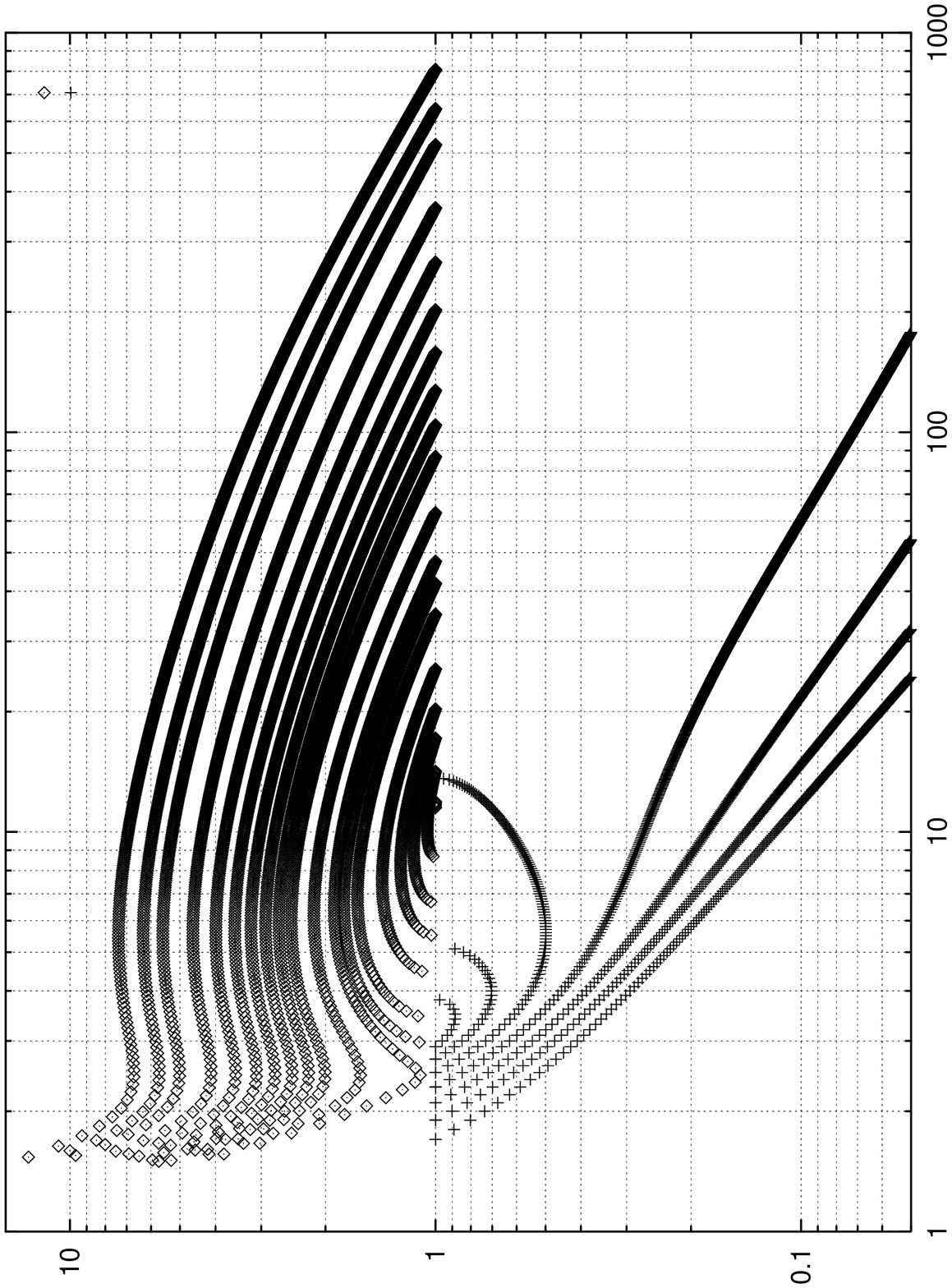}
\caption{Mach number Vs radial distance of X plA - Alp X topology for accretion rate=1.0 and $\lambda$=1.8 \label{fig2}}  
\end{figure}

\begin{figure}
\plotone{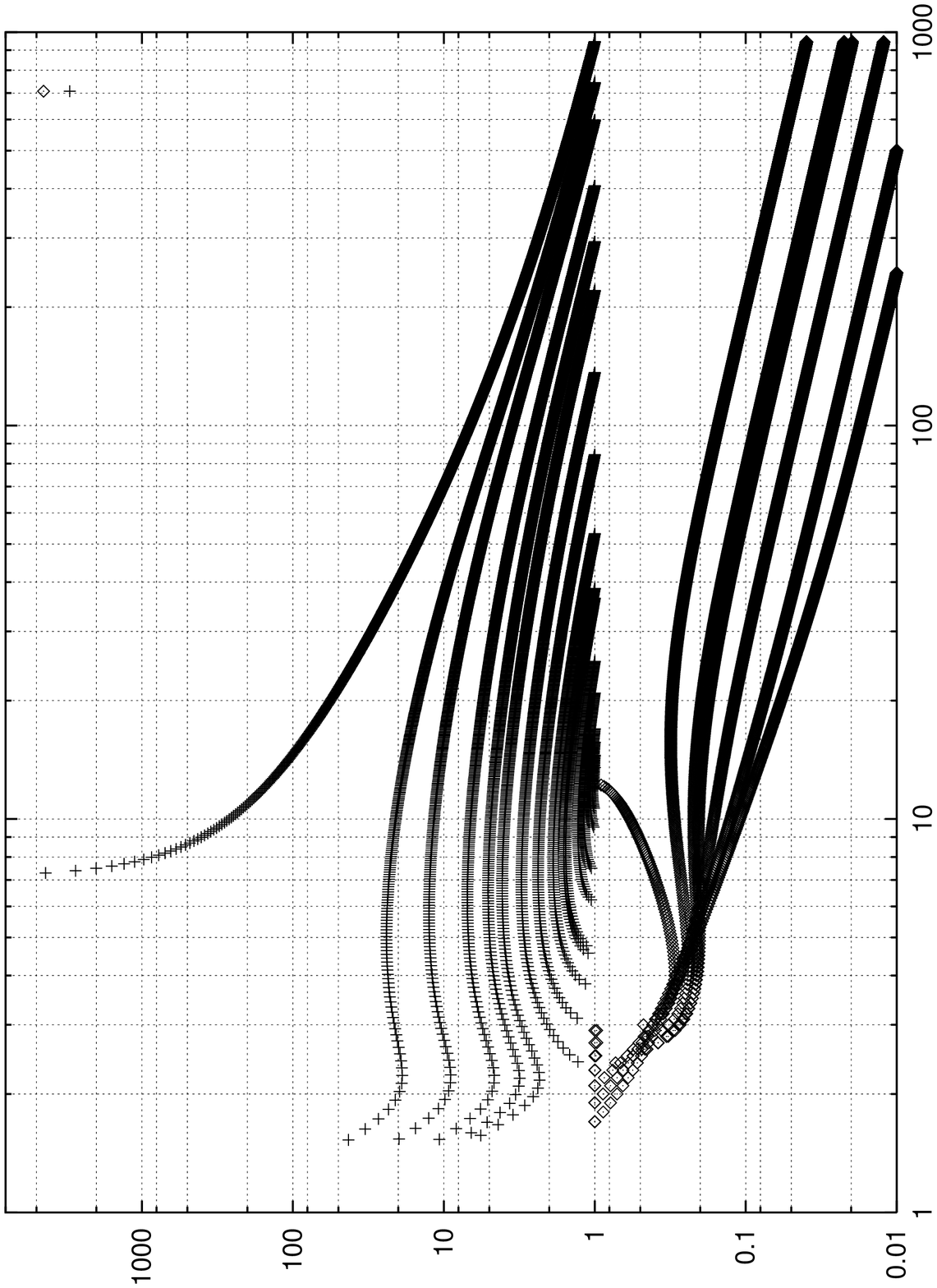} 
\caption{Mach number Vs radial distance of X lA - Alp X x topology for accretion rate=5.0 and $\lambda$=1.9 \label{fig3}}
\end{figure}

\begin{deluxetable}{clcc}
\tablehead{
\colhead{accretion rate} & \colhead{$\lambda$} & \colhead{topology} & \colhead{shock}
}

\startdata

.0001 & 1.6 & X  &  \\

.0001 & 1.65 &  X plA - Alp X &  noshock \\

.0001  &  1.7  &  X plA - Alp X &   shk  \\

.0001 & 1.9  &  X plA - Alp X & shk2 \\

.0001 & 2.0  &  X - Alp & noshock  \\  \hline

.001 & 1.6  &   X    & \\ 

.001 &  1.65 &   X plA - Alp X &  shk \\ 

.001 & 1.9 &    X plA - Alp X & shk2  \\

.001 & 2.0 &    X - Alp & noshock  \\ \hline

.01 & 1.6 &     X  & \\

.01 &  1.65 &    X plA - Alp X & shk \\ 

.01 &  1.9 &  X plA - Alp X &  shk2  \\ 

.01 &  2.0   &   X lA - Alp & noshock \\ \hline

.1  &  1.6 &      X  & \\

.1 &  1.65  &      X plA - Alp X  & shk \\

.1 &  1.9    &   X plA - Alp X & shk \\

.1  & 2.0    &   X - Alp  & noshock \\ \hline

1.0 & 1.6  &    X   & \\

1.0 & 1.65 &    X plA - Alp X & shk \\

1.0 & 1.7  &    X plA - Alp X & \\ 

1.0 & 1.8  &    X plA - Alp X & \\ 

1.0 & 1.9  &    X - Alp X  & shk2  \\ 

1.0 &  2.0 &     X - Alp   & noshock \\ 

1.0 & 2.1  &    X - Alp  & \\

1.0 & 2.3  &    X lA - Alp & \\ 

1.0 & 2.4  &    lA - Alp  & \\

1.0 & 2.5  &    lA - Alp  & \\ \hline

2.0 & 1.6 & X & \\

2.0 & 1.65  &   X plA - Alp X  & shk \\

2.0 & 1.8   &   X plA - Alp X  & \\ 

2.0 &  1.9  &    X lA - Alp X  & shk1 \\

2.0 & 2.0    &   X lA - Alp  &  noshock \\

2.0 &  2.3    &   X lA - Alp & \\ 

2.0 &  2.4      &  x lA - Alp & \\  \hline

5.0  & 1.6 &    X  &  \\

5.0  & 1.65 &    X plA - Alp X x  & shk \\

5.0  & 1.7  &    X plA - Alp X  & \\ 

5.0  &  1.8 &     X plA - Alp X & \\

5.0 &  1.9  &    X lA - Alp X x &  shk \\ 

5.0 & 2.0   &   X lA - Alp X x & noshock  \\ 

5.0 & 2.3   &   X x lA - Alp x  & \\ \hline

20. & 1.65   &  X plA - Alp X x & \\

20. &  2.4   &   X x lA - Alp x & \\  \hline

30. & 1.65  &   X plA - Alp X x & \\  \hline

50. & 1.65  &    X x   & \\

50. & 1.7  &  X plA - Alp X x &  shk \\ 

50. & 1.8    &  X plA - Alp X x  & shk1 \\

50. &  1.9  &  X plA lA - Alp X x  &  shk \\ 

50. &  2.0   &   X x lA - Alp X x  & noshock \\

50.  & 2.4   &   X x lA - Alp x  & \\  \hline

500. &  1.6 &    X x  & \\

500. &  1.8 &    X x & \\ 

500. &  2.4 &    x  & \\  \hline

\enddata

\end{deluxetable}


\newpage

\begin{thebibliography}{}

\bibitem[]{} Chakrabarti, S. K. 1989, \apj, 347, 365

\bibitem[]{} Chakrabarti, S. K. 1996, Physics Reports, 266, No. 5 \& 6, 229 

\bibitem[]{} Chakrabarti, S. K. \& Manickam, S. G. 2000 \apj, 531, L41

\bibitem[]{} Chakrabarti, S. K. \& Titarchuk, L. 1995 \apj, 455, 623

\bibitem[]{} Landau, L. D. \& Lifshitz, E. M. 1984, Fluid Mechanics, second edition, Maxwell 
Macmillan International Editions 

\bibitem[]{} Lang, K. R. 1980, Astrophysical Formulae, second edition, Springer-Verlag 

\bibitem[]{} Molteni, D., Sponholz, H. \& Chakrabarti, S. K. 1996, \apj, 457, 805, (MSC96)

\bibitem[]{} Paczynski, B. \& Wiita, P. J. 1980, A\&A, 88, 23

\bibitem[]{} Press, W. H., Teukolsky, S. A., Vellerling, W. T. \& Flannery, B. P. 1992, 
Numerical Recipes in Fortran - The Art of Scientific Computing, second edition,
Cambridge University Press 

\end{thebibliography}
\end{document}